\documentclass[fleqn,10pt]{wlscirep}
\title{Basic protocols in quantum reinforcement learning with superconducting circuits}

\author[1,*]{Lucas Lamata}
\affil[1]{Department of Physical Chemistry, University of the Basque Country UPV/EHU, Apartado  644, 48080 Bilbao, Spain}

\affil[*]{lucas.lamata@gmail.com}


\begin{abstract}
Superconducting circuit technologies have recently achieved quantum protocols involving closed feedback loops. Quantum artificial intelligence and quantum machine learning are emerging fields inside quantum technologies which may enable quantum devices to acquire information from the outer world and improve themselves via a learning process. Here we propose the implementation of basic protocols in quantum reinforcement learning, with superconducting circuits employing feedback-loop control. We introduce diverse scenarios for proof-of-principle experiments with state-of-the-art superconducting circuit technologies and analyze their feasibility in presence of imperfections. The field of quantum artificial intelligence implemented with superconducting circuits paves the way for enhanced quantum control and quantum computation protocols.\end{abstract}
\begin{document}

\flushbottom
\maketitle
\thispagestyle{empty}

\section*{Introduction}
\begin{figure}[h]
\centering
\includegraphics[width=0.8\linewidth]{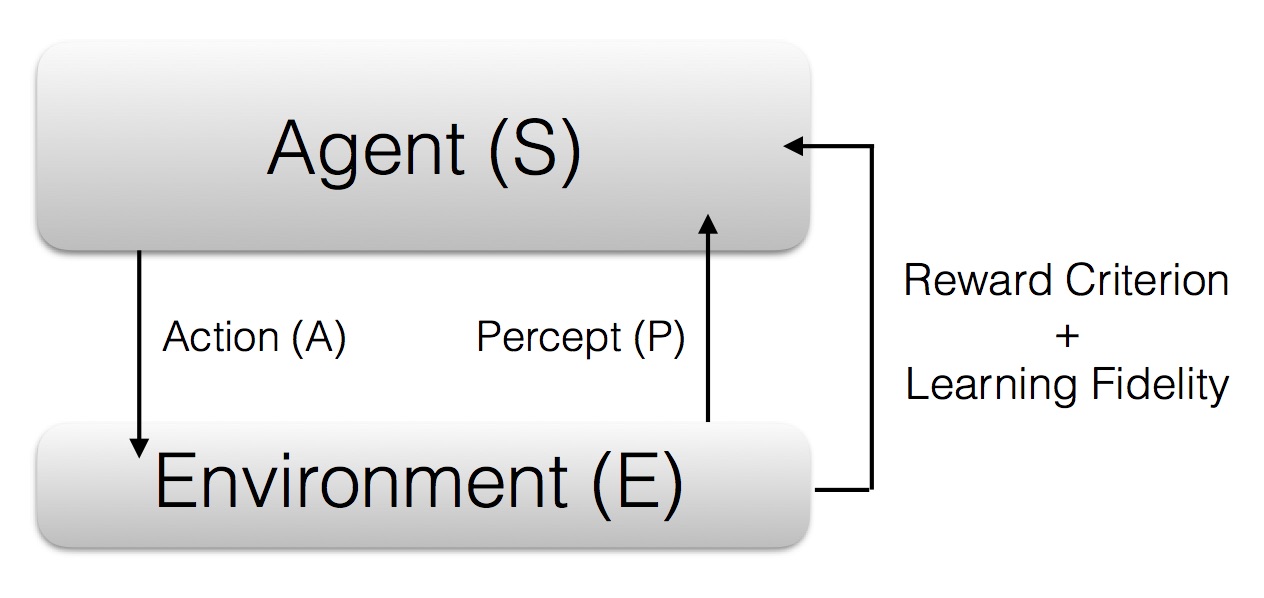}
\caption{\textbf{Scheme of reinforcement learning.} In each learning cycle, an Agent, denoted by S, interacts with an Environment, denoted by E, realizing some Action (A) on it, as well as gathering information, or Percept (P) about its relation to it. Subsequently, the information obtained is employed in order to decide a strategy on how to optimize the agent, based on a Reward Criterion, whose aim may be to maximize a Learning Fidelity. Afterwards, a new cycle begins. The situation in the quantum realm is similar, and can oscillate between having a quantum version of agent, of environment, or of both of them, as well as interactions between them that can be quantum and/or classical channels with feedforward.}\label{fig:FigRLScheme}
\end{figure}

Artificial intelligence (AI) and, inside of it, machine learning (ML), are two of the most promising research avenues in computer science nowadays~\cite{RusellAI}. ML deals with establishing dynamical algorithms with which computers can learn by themselves, without simply obeying a fixed sequence of instructions. Therefore, these devices can acquire information from the outer world, say, the ``environment'', and adapt to it, improving themselves according to some predefined criteria. The main types of ML algorithms can be classified in three varieties, namely, supervised learning, unsupervised learning, and reinforcement learning~\cite{RusellAI}. In the first case, the computer is presented with a series of classified data, which is employed in order to train the device. Later on, when unclassified data is introduced, the machine may be able to classify it, assuming that the training phase was successful. Neural networks are commonly employed for this task. In unsupervised learning, no training data is presented, but the aim is to find correlations in the available data, establishing a clustering in different groups that may be employed to classify subsequent information. Finally, in reinforcement learning, see Fig.~~\ref{fig:FigRLScheme}, perhaps the most similar ML protocol to the way the human brain works, a system (the ``agent'')  interacts with an environment, realizing some action on it, as well as gathering information about its relation to it~\cite{RLBook}. Subsequently, the information the agent obtains is employed in order to decide a strategy on how to optimize itself, based on a reward criterion, whose aim may be to maximize a learning fidelity, and afterwards the cycle starts again. 

The field of quantum technologies promises to introduce significant advantages in processing and transmission of information with respect to classical computers~\cite{NielsenChuang}. In recent years, an interest has appeared for connecting the areas of artificial intelligence, and more specifically machine learning, with quantum technologies, through the field of quantum machine learning (e.g., Refs.~\cite{PetruccioneReview,qmlLloydReview} and references therein). A specific topic that has emerged is that of quantum reinforcement learning~\cite{Dong,BriegelPRX,BriegelPRL,BoltzmannRL}. In this respect, there are promising results that point to the fact that active learning agents may perform better in a quantum scenario~\cite{BriegelPRX,BriegelPRL}. Some proposals and experimental realizations of quantum machine learning have already been performed~\cite{Sciarrino,DBrunner,JWPan,Hermans,Glaser,Zhaokai,Pons,Neven,Mabuchi,Ghosh1,Ghosh2}. On the other hand, so far the implementation of quantum reinforcement learning in current superconducting circuit technologies, including some of its prominent features as feedback-loop quantum and classical control, has not been analyzed. Related areas that have also risen in recent years include quantum artificial life~\cite{BookAL,MartinDelgado,Unai1,Unai2}, quantum memristors~\cite{Sanz,Mikel2,NoriMemristors}, as well as quantum learning based on time-delayed equations~\cite{UnaiTimeDelay1,UnaiTimeDelay2}.

Superconducting circuits are one of the leading quantum technologies in terms of quantum control, reliance, and scalability~\cite{Blais07,Wilhelm,Wendin16}. Fidelities of entangling gates larger than 99\% have been achieved on superconducting qubits~\cite{Barends14,Barends16}, and the emergence of 3D cavities has significantly improved coherence times~\cite{Paik11}. These remarkable progresses have enabled the realization of feedback-loop control in quantum protocols with superconducting circuits, which include entanglement generation via parity measurement and feedforward, as well as qubit stabilization~\cite{DiCarlo12,DevoretFeedback,DiCarlo13,DiCarlo15}. In this sense, the technology is already available in order to perform proof-of-principle experiments with simple quantum reinforcement learning protocols, to pave the way for future scalable realizations. Moreover, analyses of coherent operations in superconducting circuits that can be modified during an experiment can be also useful in this context~\cite{Friis}.

Here we propose the implementation of basic protocols in quantum reinforcement learning via feedback-loop control in state-of-the-art superconducting circuit technologies. We firstly describe elementary instances of quantum reinforcement learning algorithms, which involve all the essential ingredients of the field. These include an agent, an environment, interactions between them, projective measurements, a reward criterion, and feedforward. The motivation for considering simplified models is to make a proposal that can be carried out nowadays in the lab. Subsequently, we analyze the implementation of these minimal instances in current superconducting-circuit technologies. We point out that here we do not aim at proving quantum speedup or scalability, but to motivate experimental realizations of these building blocks to pave the way for future advances. 

\section*{Results}
\subsection*{Quantum Reinforcement Learning with Superconducting Circuits}

\begin{figure}[t]
\centering
\includegraphics[width=0.5\linewidth]{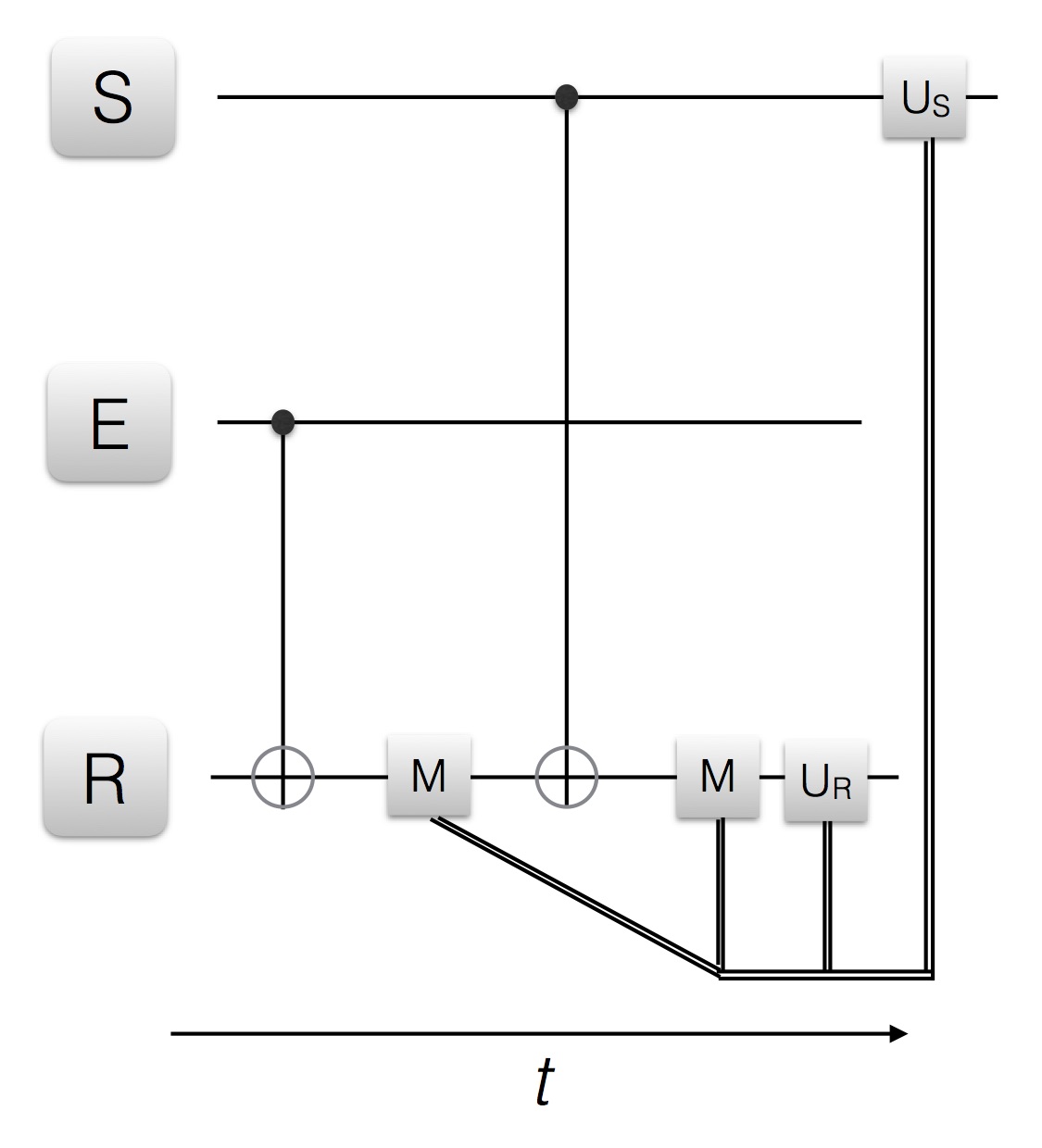}
\caption{\textbf{Quantum reinforcement learning for one qubit.} We depict the circuit representation of the proposed learning protocol. S, E and R denote the agent, environment and register qubits, respectively. CNOT gates between E and R as well as between S and R are depicted with the standard notation. M is a measurement in the computational basis chosen, while $U_S$ and $U_R$ are local operations on agent and register, respectively, conditional on the measurement outcomes via classical feedback loop. The double lines denote classical information being fedforward. The protocol can be iterated upon changes in the environment.}\label{fig:Fig1}
\end{figure}

\begin{figure}[t]
\centering
\includegraphics[width=0.5\linewidth]{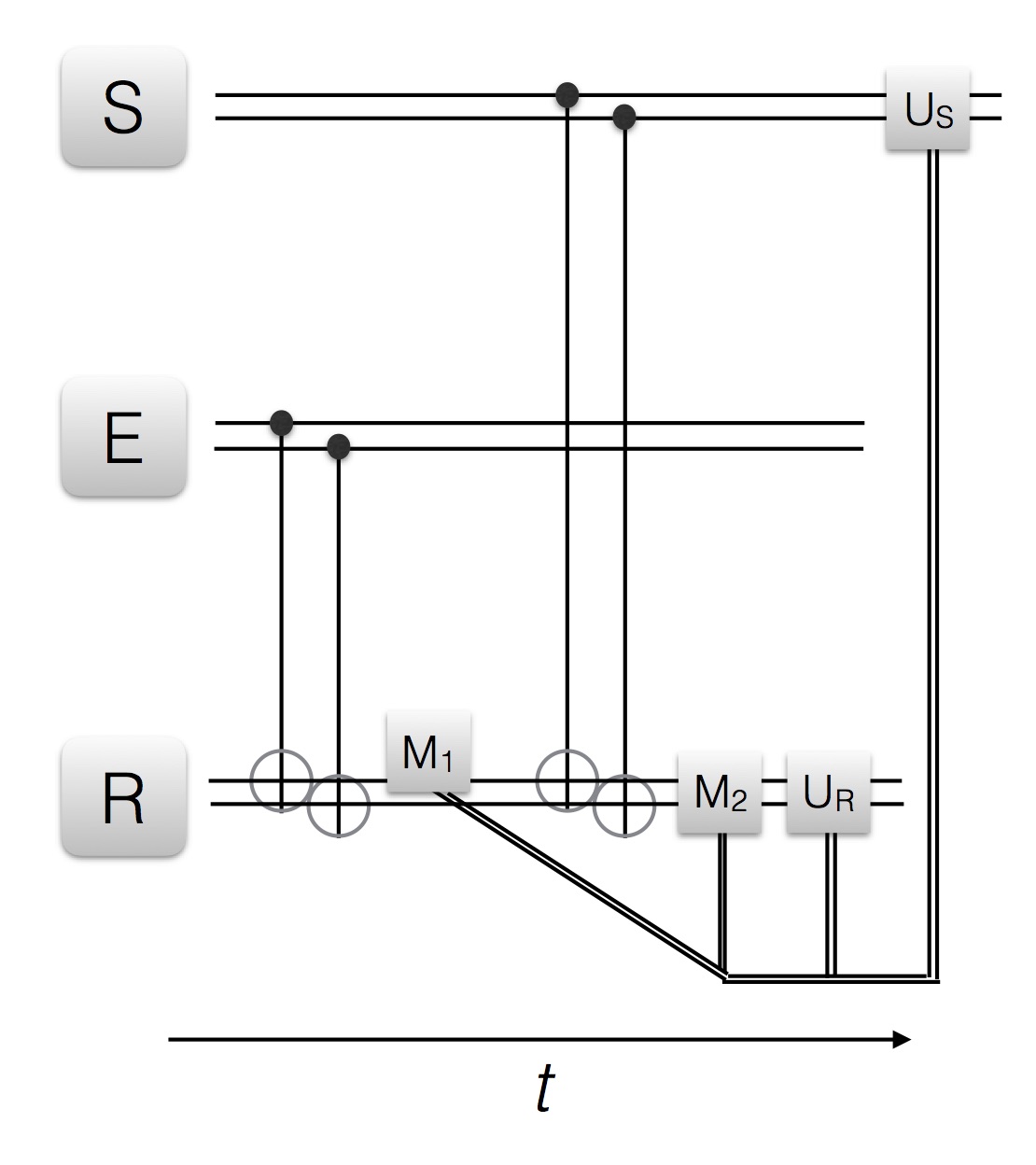}
\caption{\textbf{Quantum reinforcement learning for multiqubit system, I.} We depict the circuit representation of the proposed learning protocol. S, E and R denote the agent, environment and register two-qubit states, respectively. CNOT gates between the respective pairs of qubits in E and R as well as in S and R are depicted with the standard notation.  $M_1$ is a measurement in the computational basis chosen on the first qubit of the register, $M_2$ is a measurement in the computational basis chosen on the two-qubit state of the register, while $U_S$ and $U_R$ are local operations on agent and register, respectively, conditional on the measurement outcomes via a classical feedback loop. The double narrow lines denote classical information being fedforward, while the horizontal, double wider lines denote two-qubit states. The protocol can be iterated upon changes in the environment via reset of the agent.}\label{fig:Fig2}
\end{figure}

We firstly consider a set (agent, environment, register) composed of one qubit each, and later on we will extend it to more complex situations involving two-qubit states for each of these parts. The use of an auxiliary register qubit may prove useful in some implementations, diverting the necessary projective (or other kind of) measurements onto the auxiliary system instead of measuring directly the environment and agent qubits. Besides superconducting circuits, other quantum technologies as trapped ions and quantum photonics may benefit from this approach, avoiding ion heating during resonance fluorescence as well as photon loss due to destructive measurement. The quantum reinforcement learning protocols we envision are depicted in Figs.~\ref{fig:Fig1},\ref{fig:Fig2},\ref{fig:Fig3}. In the single-qubit case, a CNOT gate is applied on the environment-register subspace and the register is measured in order to acquire information about the environment (Action A, see Fig.~\ref{fig:FigRLScheme}). Subsequently, a CNOT gate is applied on the agent-register subspace and the register is measured, which provides information about the agent (Percept P, see Fig.~\ref{fig:FigRLScheme}). Finally, the reward criterion is applied with the information of the previous measurements, and the agent is updated accordingly via local operations mediated with a closed feedback loop, with the aim of maximizing the learning fidelity. For the reward criterion we will consider to have maximum overlap (or maximum positive correlation in the presence of entanglement, in the multiqubit case) between agent and environment. A possible motivation for this approach can be to partially clone~\cite{Unai1} quantum information from the environment onto the agent. In the examples considered, a single learning iteration suffices to achieve the maximum learning fidelity. Nevertheless, in a changing environment, the process may be continuously implemented for an optimal performance, given that the agent should adapt to each new environment in order to maximize the learning fidelity. Additionally, for more complex situations with larger agents and environments several iterations may be needed for convergence to large learning fidelities, involving complex learning curves as in standard machine learning protocols. Our approach is a simplified protocol of quantum reinforcement learning, implying that at the end of the learning process the agent has a complete description of the environment, that is, a model, whereas usually reinforcement learning schemes are model-free. Moreover, the reward function is taken for simplicity in our case to be the fidelity, although this could be modified to allow for more general situations. In our protocol, the agent acts on the environment via the initial projective measurement on the latter, which will modify the state of the environment. We point out that, in the single-qubit agent and environment case, the learning process is purely classical. This kind of quantum approach to classical learning has also been analyzed in different works~\cite{classicalLearning1,classicalLearning2}. On the other hand, in the multiqubit case also analyzed in this article, entanglement between agent and environment states is produced, and therefore the situation in this case is to have a genuine quantum scenario. The motivation for our study is to propose an implementation of basic quantum reinforcement learning protocols with current superconducting circuit technology. Thus, the complexity of the analyzed systems is kept small in order that an experimental realization may be presently carried out.

We exemplify now the proposed protocol with a set of cases, in growing complexity. We first introduce the quantum information algorithm examples and later on we analyze their feasibility with current superconducting circuit technology.

\subsubsection*{Single-qubit agent and environment states}

We introduce now some examples for single-qubit agent, environment, and register states.

i) $\{|S\rangle_0=|0\rangle,|E\rangle_0=|0\rangle,|R\rangle_0=|0\rangle\}$  

This is a trivial example in which the initial state of the agent, $|S\rangle_0$, already maximizes the overlap with the environment state, $|E\rangle_0$, such that this is a fixed point of the protocol and no additional dynamics occurs, unless the environment subsequently evolves, which will produce the adaptation of the agent.

ii) $\{|S\rangle_0=|0\rangle,|E\rangle_0=(|0\rangle+|1\rangle)/\sqrt{2},|R\rangle_0=|0\rangle\}$

The first step of the protocol consists in acquiring information from the environment and transferring it to the register, in order that, later on, the agent state can be updated accordingly conditional on the environment state, see Fig.~\ref{fig:Fig1}. Therefore, the first action is a CNOT gate on the environment-register subspace, where the environment qubit acts as the control and the register qubit acts as the target,
\begin{equation}
U_{\rm CNOT} |E\rangle_0|R\rangle_0\equiv |ER\rangle_{0\rightarrow1}=\frac{1}{\sqrt{2}}(|00\rangle+|11\rangle).
\end{equation}
Subsequently, the register qubit is measured in the $\{|0\rangle,|1\rangle\}$ basis, giving as outcomes the $|0\rangle$ or $|1\rangle$ states for the register, with 1/2 probability each. The following step is to update the agent state according to the register state, namely, for $|R\rangle=|0\rangle$, the action on the agent is the identity gate, while for $|R\rangle=|1\rangle$ it is an $X$ gate.
Therefore, in the first case, $|S\rangle_1=|E\rangle_1=|0\rangle$, while in the second case, $|S\rangle_1=|E\rangle_1=|1\rangle$, such that the reward criterion is succesfully applied, and the learning fidelity is maximal, $F_S\equiv|_1\langle E|S\rangle_1|^2=1$. Finally, the register state is updated to initialitize it back onto the $|R\rangle_0$ state.

We point out some remarks:

The state of the environment has been changed during the protocol, according to the approach here followed. In a fully quantum reinforcement learning algorithm, either agent, environment, or both, should be quantum and measured (at least partially) in each iteration of the process, collapsing their respective states\cite{Dong,BriegelPRX,BriegelPRL}. In previous results in the literature analyzing quantum reinforcement learning protocols, the emphasis was put on a quantum agent interacting with a classical or oracular quantum environment~\cite{BriegelPRX,BriegelPRL}, in which the learning was produced by the computational power of the agent or resulted from a complex algorithm with the oracle, respectively. Here we focus on simple instances of quantum agent and environment, to explore this situation and their mutual quantum interactions, while always having in mind possible implementations with current superconducting-circuit technology. As we have just shown, the environment state would change after the initial agent's measurement of it, as expected. Nevertheless, the figure of merit, the learning fidelity obtained after application of the reward criterion, is here maximal with respect to the new environment state, namely, the agent has adapted to this new environment. In larger, Markovian reservoirs, the effect of the measurement on the environment may be possibly disregarded, although in many cases the agent will collapse to the dark state of the Liouville operator without the need for a closed feedback loop. The reason for this is that, in several instances in which a system strongly couples to a reservoir, the asymptotic state of the system will be approximately given by the stationary state in which the system does not evolve further. This will correspond, in the case of purely dissipative dynamics, to the dark state of the Liouvillian. Most likely a highly interesting situation will come from mesoscopic, non-Markovian environments, which however are outside the scope of this work for not being currently straightforwardly achievable.

The projective measurement in the $\{|0\rangle,|1\rangle\}$ basis already selects it as the final basis for environment and agent according to the present protocol, namely, the final environment and agent states that are achieved are product states which are elements of this basis, at least in the single-qubit case. In order to attain other agent/environment states in a rotated $\{|\tilde{0}\rangle,|\tilde{1}\rangle\}$ basis, the projective measurements in the protocol should be carried out in this new basis. In general, for multiqubit agent and environment states, the measurement basis may also be generalized to be entangled, or partial measurements  (see below), Positive Operator Valued Measures (POVM), as well as weak measurements, may be performed. 

The quantum character in the single-qubit case comes from the fact that the agent, environment, and register states are quantum, as well as the CNOT gates applied on them. On the other hand, there is no entanglement in the projected states, and, as we just pointed out, the measurement basis representatives select the outcome basis, which could be orthonormal to the original environment state, as in case ii) above. Therefore, in this case the environment is projected onto the orthogonal complement with a totally random output. Other options are to employ POVM, or to consider measurement bases with large overlap of some representative with the environment at all times. Nevertheless, even in the case ii) above, the agent learns the environment information with certainty after the environment projection onto the measurement basis. In the multiqubit case considered below, the final agent-environment state is entangled, therefore genuinely quantum.

iii) $\{|S\rangle_0=\alpha_S|0\rangle+\beta_S|1\rangle,|E\rangle_0=\alpha_E|0\rangle+\beta_E|1\rangle,|R\rangle_0=|0\rangle\}$

The first step of the protocol in this case consists in a CNOT gate on the environment-register subspace, where, as before, the environment qubit and the register qubit are control and target, respectively,
\begin{equation}
U_{\rm CNOT} |E\rangle_0|R\rangle_0\equiv |ER\rangle_{0\rightarrow1}=\alpha_E|00\rangle+\beta_E|11\rangle,
\end{equation}
where the first qubit represents the environment and the second one the register.
Afterwards, the register qubit is measured in the $\{|0\rangle,|1\rangle\}$ basis, producing the $|0\rangle$ or $|1\rangle$ states for the register, with probabilities $|\alpha_E|^2$ for the 0 and $|\beta_E|^2$ for the 1.

The next step will be to apply a CNOT gate on the agent-register subspace, where the agent qubit and the register qubit are respectively control and target, e.g., for the 0 outcome in the previous measurement,
\begin{equation}
U_{\rm CNOT} |S\rangle_0|R\rangle_0\equiv |SR\rangle_{0\rightarrow1}=\alpha_S|00\rangle+\beta_S|11\rangle,
\end{equation}
where the first subspace represents the agent and the second the register.
Subsequently, the register qubit is measured in the same basis as before, giving as outcomes the $|0\rangle$ or $|1\rangle$ states, with probabilities $|\alpha_S|^2$ for the 0 and $|\beta_S|^2$ for the 1.

Followingly, we update the agent and register states according to the two register state measurements, namely, for $|R\rangle_{M_E,M_S}=|00\rangle$, the action on agent and register is the identity gate. The probability for this case to happen is $|\alpha_E|^2|\alpha_S|^2$. For $|R\rangle_{M_E,M_S}=|01\rangle$, the action on agent and register is the $X$ gate for both. The probability for this case is $|\alpha_E|^2|\beta_S|^2$. For $|R\rangle_{M_E,M_S}=|10\rangle$, the action on agent and register is the identity gate. The probability for this case is $|\beta_E|^2|\beta_S|^2$. Finally, for $|R\rangle_{M_E,M_S}=|11\rangle$, the action on agent and register is the $X$ gate. The probability for this case is $|\beta_E|^2|\alpha_S|^2$. Here, $|R\rangle_{M_E,M_S}$ corresponds to the pair of register states, each obtained after each measurement, being $M_E$ the corresponding one after interaction with the environment, and $M_S$ the corresponding one after interaction with the agent.

Therefore, in the first two cases, $|S\rangle_1=|E\rangle_1=|0\rangle$, while in the second two cases, $|S\rangle_1=|E\rangle_1=|1\rangle$, such that the reward criterion is succesfully applied, and the learning fidelity is maximal, $F_S\equiv|_1\langle E|S\rangle_1|^2=1$. Moreover, this way the register is at the end initialized back onto the $|R\rangle_0$ state.

Subsequently, the environment state may change, and the protocol should be run again in order that the state adapts to these changes. For this simplified single-qubit agent model, a single learning iteration is enough in order to achieve maximum learning fidelity for given agent and environment initial states. In more complex multiqubit agents, and employing partial measurements or weak measurements, further iterations may be needed in order to maximize the fidelity. We also point out that the case in which the initial environment state has a large overlap with one of the measurement basis states will increase the fidelity  $|_0\langle E|S\rangle_1|^2$ between this environment state and the achieved agent state after the learning protocol. Nevertheless, we remark that in this model, and for any initial environment state, the agent learns the final environment state, $|E\rangle_1$, with certainty, i.e., deterministically and with learning fidelity 1.

\subsubsection*{Multiqubit agent and environment states}

We give now some examples for multiqubit agent, environment, and register states.

i) $\{|S\rangle_0=\alpha^{00}_S|00\rangle+\alpha_S^{01}|01\rangle+\alpha^{10}_S|10\rangle+\alpha_S^{11}|11\rangle,|E\rangle_0=\alpha^{00}_E|00\rangle+\alpha_E^{01}|01\rangle+\alpha^{10}_E|10\rangle+\alpha_E^{11}|11\rangle,|R\rangle_0=|00\rangle\}$

We consider now a multiqubit case and, focusing on projective measurements as before, one can choose between measuring both register qubits or only one of them. Measuring both will project agent and environment states onto some among the measurement basis representatives, while measuring only one register qubit may preserve part of the agent-environment entanglement, adding further complexity and quantumness to the analysis. We will describe each of these cases in the following.
   
   The first step of the protocol in this multiqubit case consists in applying two CNOT gates on the environment-register subspace, where the $k$th environment qubit acts as the control and the $k$th register qubit acts as the target in the $k$th CNOT gate, $k=1,2$.
\begin{equation}
U_{{\rm CNOT},1} U_{{\rm CNOT},2}|E\rangle_0|R\rangle_0\equiv |ER\rangle_{0\rightarrow1}=\alpha_E^{00}|00\rangle_E|00\rangle_R+\alpha_E^{01}|01\rangle_E|01\rangle_R+\alpha_E^{10}|10\rangle_E|10\rangle_R+\alpha_E^{11}|11\rangle_E|11\rangle_R,\label{MultiqubitER}
\end{equation}
where the first two qubits represent the environment and the second ones the register.

\begin{figure}[t]
\centering
\includegraphics[width=0.5\linewidth]{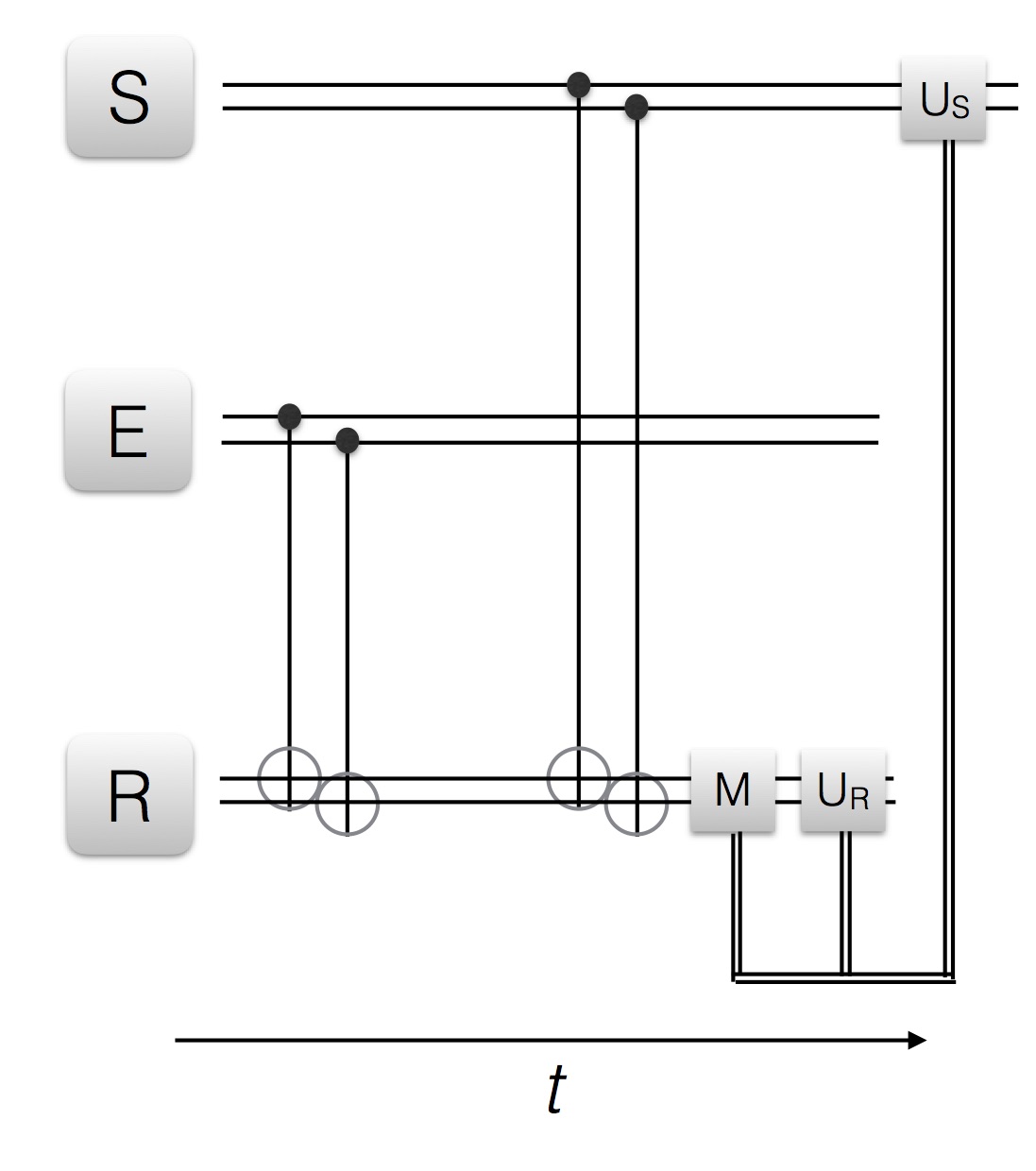}
\caption{\textbf{Quantum reinforcement learning for multiqubit system, II.} We depict the circuit representation of the proposed learning protocol. S, E and R denote the agent, environment and register two-qubit states, respectively. CNOT gates between the respective pairs of qubits in E and R as well as in S and R are depicted with the standard notation.  In this case, the measurement $M$ acts on both register qubits, and is performed in the computational basis chosen. $U_S$ and $U_R$ are local operations on agent and register, respectively, conditional on the measurement outcomes via a classical feedback loop. The double narrow lines denote classical information being fedforward, while the horizontal, double wider lines denote two-qubit states. The protocol can be iterated upon changes in the environment via reset of the agent.}\label{fig:Fig3}
\end{figure}

We now differentiate the cases in which i) we perform a complete two-qubit measurement on the register, which will provide final agent and environment product states. These will belong to the representatives of the measurement basis, as in the single-qubit case. The maximization of the learning fidelity will be given, as before, by the environment-agent state overlap, which can be achieved deterministically and with fidelity 1. ii) we perform a single-qubit measurement on one of the register qubits, after interaction with the environment, see Fig.~\ref{fig:Fig2}. This will have as a consequence the preservation of part of the agent-environment entanglement at the end of the protocol. The maximization of the learning fidelity will be given in this case by attaining a maximally correlated agent-environment state, such that any local measurement in the considered basis carried out in the agent will give the same outcome for the environment, and viceversa. In the entangled-state context, and for the chosen measurement basis, this criterion will assure that agent and environment will behave identically under measurements such that the agent will have learned the environment structure, at the same time as modifying it by entangling to it. iii) we perform no measurement after the environment-register interaction and before the agent-register interaction, see Fig.~\ref{fig:Fig3}. Therefore, a larger entanglement between agent and environment will now be achieved and, equivalently to the case ii), the rewarding criterion will be based on attaining maximum positively-correlated agent-environment states.

{\it i) Total two-qubit measurement and product final states.--} 

Subsequently, the register qubits are measured in the $\{|00\rangle,|01\rangle,|10\rangle,|11\rangle\}$ basis, giving as outcomes the basis states for register and environment, with probabilities $|\alpha_E^{ij}|^2$.

  The next step consists in applying two CNOT gates on the agent-register subspace, where the $k$th agent qubit acts as the control and the $k$th register qubit acts as the target in the $k$th CNOT gate, $k=1,2$. We consider, without loss of generality, that the outcome in the first measurement was $|R\rangle_{M_E}=|00\rangle$, with probability $|\alpha_E^{00}|^2$. The other three cases can be computed similarly to this one. We thus have,
\begin{equation}
U_{{\rm CNOT},1} U_{{\rm CNOT},2}|S\rangle_0|R\rangle_0\equiv |SR\rangle_{0\rightarrow1}=\alpha_S^{00}|00\rangle_S|00\rangle_R+\alpha_S^{01}|01\rangle_S|01\rangle_R+\alpha_S^{10}|10\rangle_S|10\rangle_R+\alpha_S^{11}|11\rangle_S|11\rangle_R,
\end{equation}
where the first two qubits represent the agent and the second ones the register. Subsequently, the register qubits are measured in the $\{|00\rangle,|01\rangle,|10\rangle,|11\rangle\}$ basis, giving as outcomes the basis states for register and agent, with probabilities $|\alpha_S^{ij}|^2$.

Followingly, we update the agent and register states according to the two register state measurements, namely, for $|R\rangle_{M_E,M_S}=|00,00\rangle$, the action on the state and register is the two-qubit identity gate. The probability for this case to happen is $|\alpha_E^{00}|^2|\alpha_S^{00}|^2$. For $|R\rangle_{M_E,M_S}=|00,01\rangle$, the action on the state and register is the $I\otimes X$ gate for each. The probability for this case is $|\alpha_E^{00}|^2|\alpha_S^{01}|^2$. The other cases can be similarly computed. Here, $|R\rangle_{M_E,M_S}$ corresponds to the pair of two-qubit register states, each obtained after each measurement, being $M_E$ the corresponding one after interaction with the environment, and $M_S$ the corresponding one after interaction with the agent.

Therefore, in the case with $|R\rangle_{M_E,M_S}=|ij,kl\rangle$, $|S\rangle_1=|E\rangle_1=|ij\rangle$, $i,j=0,1$, such that the reward criterion is succesfully applied, and the learning fidelity is maximal, $F_S\equiv|_1\langle E|S\rangle_1|^2=1$. This is achieved, after the feedback-loop application, deterministically and with fidelity 1. The register qubits are as well initialized back onto the $|R\rangle_0=|00\rangle$ state.

{\it ii) Partial, single-qubit measurement and entangled agent-environment final state.--} 

We consider an alternative case in which, instead of measuring the two register qubits after interaction with the environment, we only observe the first one, see Fig.~\ref{fig:Fig2}. Therefore, beginning with Eq.~(\ref{MultiqubitER}),  we propose to measure the first qubit of the register state, in the same basis as before. Assuming that coherence between compatible results can be maintained during the projection, as happens, e.g., in current realizations of feedback-controlled experiments with superconducting circuits~\cite{DiCarlo13,DiCarlo15}, we obtain, for the 0 outcome, achieved with probability $|\alpha_E^{00}|^2+|\alpha_E^{01}|^2$,
\begin{equation}
|ER\rangle_{0\rightarrow1}\propto\alpha_E^{00}|00\rangle_E|00\rangle_R+\alpha_E^{01}|01\rangle_E|01\rangle_R,\label{MeasureERMulti}
\end{equation}
while the 1 outcome can be similarly computed.

Focusing on the previous 0 outcome case, we now apply the subsequent part of the protocol, namely, two CNOT gates on the agent-register subspace, where the $k$th agent qubit acts as the control and the $k$th register qubit acts as the target in the $k$th CNOT gate, $k=1,2$. Here an entangled agent-environment-register state will be obtained. We thus have,
\begin{eqnarray}\nonumber
U_{{\rm CNOT},1} U_{{\rm CNOT},2}|S\rangle_0|ER\rangle_{0\rightarrow 1} & \equiv & |SER\rangle_{0\rightarrow1}\propto \alpha_E^{00}(\alpha_S^{00}|00\rangle_S|00\rangle_R+\alpha_S^{01}|01\rangle_S|01\rangle_R+\alpha_S^{10}|10\rangle_S|10\rangle_R+\alpha_S^{11}|11\rangle_S|11\rangle_R)|00\rangle_E\\&&+\alpha_E^{01}(\alpha_S^{00}|00\rangle_S|01\rangle_R+\alpha_S^{01}|01\rangle_S|00\rangle_R+\alpha_S^{10}|10\rangle_S|11\rangle_R+\alpha_S^{11}|11\rangle_S|10\rangle_R)|01\rangle_E.
\end{eqnarray}
Subsequently, the register qubits are measured in the $\{|00\rangle,|01\rangle,|10\rangle,|11\rangle\}$ basis, giving as outcomes $|SE\rangle^{ij}_1$: 
\begin{eqnarray}
|R\rangle_{M_S} & = & |00\rangle\rightarrow |SE\rangle^{00}_1\propto\alpha_S^{00}\alpha_E^{00}|00\rangle_S|00\rangle_E+\alpha_S^{01}\alpha_E^{01}|01\rangle_S|01\rangle_E,\nonumber\\
|R\rangle_{M_S} & = & |01\rangle\rightarrow |SE\rangle^{01}_1\propto \alpha_S^{01}\alpha_E^{00}|01\rangle_S|00\rangle_E+\alpha_S^{00}\alpha_E^{01}|00\rangle_S|01\rangle_E,\nonumber\\
|R\rangle_{M_S} & = & |10\rangle\rightarrow |SE\rangle^{10}_1\propto\alpha_S^{10}\alpha_E^{00}|10\rangle_S|00\rangle_E+\alpha_S^{11}\alpha_E^{01}|11\rangle_S|01\rangle_E,\nonumber\\
|R\rangle_{M_S} & = & |11\rangle\rightarrow |SE\rangle^{11}_1\propto\alpha_S^{11}\alpha_E^{00}|11\rangle_S|00\rangle_E+\alpha_S^{10}\alpha_E^{01}|10\rangle_S|01\rangle_E.
\end{eqnarray}
Followingly, in order to obtain maximal positively-correlated states, we update the agent qubits according to the two-qubit register state measurements, namely, for $|R\rangle_{M_S}=|00\rangle$, the action on the agent state is the two-qubit identity gate. For $|R\rangle_{M_S}=|01\rangle$, the action on the agent state is the $I\otimes X$ gate. For $|R\rangle_{M_S}=|10\rangle$, the action on the agent state is the $X\otimes I$ gate. Finally, for the $|R\rangle_{M_S}=|11\rangle$, the action on the agent state is the $X\otimes X$ gate. The outcome in all cases is, deterministically and with fidelity 1, a maximal positively-correlated agent-environment state, in the sense that any local measurement in the considered basis will give the same outcomes for agent and environment. Therefore, one can consider that the agent, by entangling with the environment, has achieved to learn its structure. Notice that in all cases the agent and environment final state components correspond to the ones of the environment after the initial single-qubit register measurement, see Eq.~(\ref{MeasureERMulti}). Thus, the environment structure after this measurement is preserved after the feedback loop onto the agent state, except that the agent is now entangled with the environment duplicating this structure with the corresponding weights, which contain information about the initial agent state. For an initial agent state with all four amplitudes equal, the final agent-environment entangled state will coincide with the state in Eq.~(\ref{MeasureERMulti}) via substituting the register state by the agent state. We point out that this case, containing entanglement in the final agent-environment state, has a fully quantum character. Given this presence of quantum correlations, the figure of merit, i.e., the learning fidelity, must be modified in order to make it compatible with the new situation. This is the reason why we consider in this case the achievement of maximal positively-correlated agent-environment states as the signature for maximizing the learning fidelity. The register qubits are also at the end initialized back onto the $|R\rangle_0=|00\rangle$ state.

{\it iii) Case without measurement after environment-register interaction and entangled agent-environment final state.--} 

In this third case, instead of performing a measurement on the register after interaction with the environment, we couple the environment-register subspace with the agent already, see Fig.~\ref{fig:Fig3}. Therefore, equivalently to Eq.~(\ref{MultiqubitER}), the next step consists of two CNOT gates on agent and register, where the $k$th agent qubit is the control and the $k$th register qubit is the target in the $k$th CNOT gate, $k=1,2$. In this case we will also achieve an entangled agent-environment-register state. We thus have,
\begin{eqnarray}\nonumber
&&(U_{{\rm CNOT},1} U_{{\rm CNOT},2})_{SR}(U_{{\rm CNOT},1} U_{{\rm CNOT},2})_{ER}|S\rangle_0|E\rangle_0|R\rangle_{0}  \equiv  |SER\rangle_{0\rightarrow1}\propto\\&& \alpha_E^{00}(\alpha_S^{00}|00\rangle_S|00\rangle_R+\alpha_S^{01}|01\rangle_S|01\rangle_R+\alpha_S^{10}|10\rangle_S|10\rangle_R+\alpha_S^{11}|11\rangle_S|11\rangle_R)|00\rangle_E\\\nonumber&&+\alpha_E^{01}(\alpha_S^{00}|00\rangle_S|01\rangle_R+\alpha_S^{01}|01\rangle_S|00\rangle_R+\alpha_S^{10}|10\rangle_S|11\rangle_R+\alpha_S^{11}|11\rangle_S|10\rangle_R)|01\rangle_E\\\nonumber &&+\alpha_E^{10}(\alpha_S^{00}|00\rangle_S|10\rangle_R+\alpha_S^{01}|01\rangle_S|11\rangle_R+\alpha_S^{10}|10\rangle_S|00\rangle_R+\alpha_S^{11}|11\rangle_S|01\rangle_R)|10\rangle_E\\\nonumber &&+\alpha_E^{11}(\alpha_S^{00}|00\rangle_S|11\rangle_R+\alpha_S^{01}|01\rangle_S|10\rangle_R+\alpha_S^{10}|10\rangle_S|01\rangle_R+\alpha_S^{11}|11\rangle_S|00\rangle_R)|11\rangle_E.
\end{eqnarray}
Afterwards, the register qubits should be measured in the $\{|00\rangle,|01\rangle,|10\rangle,|11\rangle\}$ basis, giving as outcomes $|SE\rangle^{ij}_1$: 
\begin{eqnarray}
|R\rangle_{M_S} & = & |00\rangle\rightarrow |SE\rangle^{00}_1\propto\alpha_S^{00}\alpha_E^{00}|00\rangle_S|00\rangle_E+\alpha_S^{01}\alpha_E^{01}|01\rangle_S|01\rangle_E+\alpha_S^{10}\alpha_E^{10}|10\rangle_S|10\rangle_E+\alpha_S^{11}\alpha_E^{11}|11\rangle_S|11\rangle_E,\nonumber\\
|R\rangle_{M_S} & = & |01\rangle\rightarrow |SE\rangle^{01}_1\propto \alpha_S^{01}\alpha_E^{00}|01\rangle_S|00\rangle_E+\alpha_S^{00}\alpha_E^{01}|00\rangle_S|01\rangle_E+\alpha_S^{11}\alpha_E^{10}|11\rangle_S|10\rangle_E+\alpha_S^{10}\alpha_E^{11}|10\rangle_S|11\rangle_E,\nonumber\\
|R\rangle_{M_S} & = & |10\rangle\rightarrow |SE\rangle^{10}_1\propto\alpha_S^{10}\alpha_E^{00}|10\rangle_S|00\rangle_E+\alpha_S^{11}\alpha_E^{01}|11\rangle_S|01\rangle_E+\alpha_S^{00}\alpha_E^{10}|00\rangle_S|10\rangle_E+\alpha_S^{01}\alpha_E^{11}|01\rangle_S|11\rangle_E,\nonumber\\
|R\rangle_{M_S} & = & |11\rangle\rightarrow |SE\rangle^{11}_1\propto\alpha_S^{11}\alpha_E^{00}|11\rangle_S|00\rangle_E+\alpha_S^{10}\alpha_E^{01}|10\rangle_S|01\rangle_E+\alpha_S^{01}\alpha_E^{10}|01\rangle_S|10\rangle_E+\alpha_S^{00}\alpha_E^{11}|00\rangle_S|11\rangle_E.\label{FinalQRLstateMulti}
\end{eqnarray}
Subsequently, in order to achieve the maximal positively-correlated states, one will now act on the agent qubits according to the two-qubit register state measurement outcomes, i.e., for $|R\rangle_{M_S}=|00\rangle$, the action on the agent state is the two-qubit identity gate. For $|R\rangle_{M_S}=|01\rangle$, the action on the agent state is the $I\otimes X$ gate. For $|R\rangle_{M_S}=|10\rangle$, the action on the agent state is the $X\otimes I$ gate. Finally, for the $|R\rangle_{M_S}=|11\rangle$, the action on the agent state is the $X\otimes X$ gate. As in previous case ii), the outcome is, deterministically and with fidelity 1, a maximal positively-correlated agent-environment state, namely, with equal outcomes for local measurements of agent and environment in the chosen basis. Moreover, for the case in which the initial agent state has all four amplitudes equal, the final agent-environment state is identical to the initial environment state, with the substitutions $|i,j\rangle_E\rightarrow|i,j\rangle_S|i,j\rangle_E$. Therefore, not only the agent and environment are maximally positively correlated, but also the amplitudes of the entangled state reproduce the ones of the initial state of the environment, namely, the quantum information of the environment has been transferred to the collective agent-environment state. Even though the final state in this case could be also achieved via initialization of the agent state in the $|00\rangle_S$ state and applying two CNOT gates to environment and agent, we point out that the outcomes in Eqs.~(\ref{FinalQRLstateMulti}) are general, valid for any two-qubit initial agent and environment states. Similarly as before, the register two-qubit state is also at the end initialized back onto the $|R\rangle_0=|00\rangle$ state.

Upon a changing environment, the agent state should be disentangled from it via reset, which can be also done via feedback~\cite{DiCarlo15} and the protocol started again in order to adapt to the new situation. The initial agent and environment states we consider in the multiqubit case exposed above in this section are general two-qubit pure states, such that the agent can begin in any state and subsequently learn any new configuration of the environment.
 
\subsection*{Analysis of a possible implementation with current superconducting circuit technology}

\begin{figure}[t]
\centering
\includegraphics[width=0.6\linewidth]{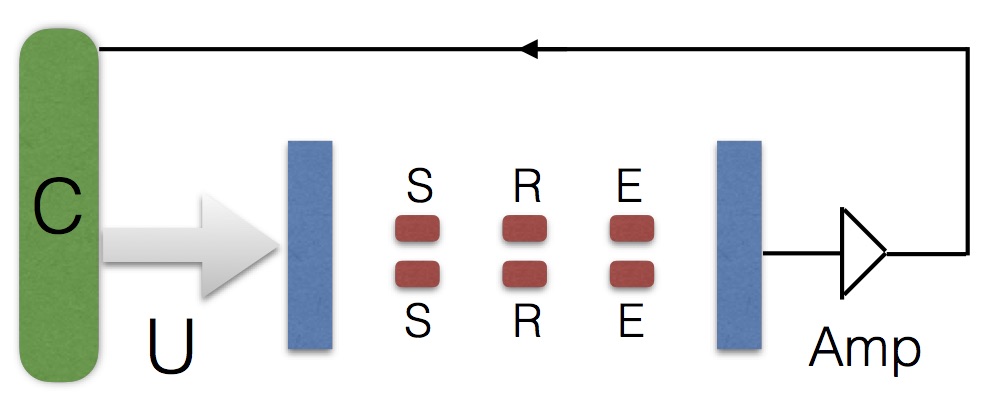}
\caption{\textbf{Scheme of the proposed implementation.}  In the most complex example proposed, we consider 6 superconducting qubits inside a 3D cavity, distributed in two rows along the cavity axis (another possible configuration would be with two 3-qubit columns perpendicular to the cavity axis). Amp denotes the amplification process, while C represents the controller device, and $U$ is a local operation on the qubits conditional on the classical feedback loop.}\label{fig:Fig4}
\end{figure}

The ingredients for the efficient implementation of these basic protocols of quantum reinforcement learning with superconducting circuits are already technologically available. Basically, one requires i) a long coherence time of the qubits, ii) high-fidelity two-qubit gates, iii) a high-fidelity projective readout, and iv) fast closed-loop feedback control conditional on the measurement outcome. We now overview these issues in the light of our protocols and considering current achievable parameters. Our proposed implementation is schematized in Fig.~\ref{fig:Fig4}.

\subsubsection*{Long coherence times}

The appearance of 3D cavities in superconducting circuits~\cite{Paik11} has extended the coherence times of qubits to $> 10 \mu$s. Other kinds of qubits, as Xmons, have also significant coherence times~\cite{Barends14}. Therefore, more than 1000 Rabi oscillations can be performed nowadays even in absence of active error correction~\cite{Barends14}. The simplicity of the quantum circuits proposed here makes them fully feasible to be carried out inside the coherence times of the qubits.

\subsubsection*{High-fidelity two-qubit gates}

Single-qubit and two-qubit gate fidelities have also experienced a significant boost in recent years, with numbers already outperforming 99.9\% for single-qubit and 99\% for two-qubit operations, for example, with Xmons via capacitive coupling~\cite{Barends14,Barends16}. Other kinds of two-qubit gates, e.g., via a resonator-mediated quantum bus, have also large fidelities, approaching 99\%. In all our protocols the number of CNOT gates that are employed is four or less. Therefore, for an appropriate qubit configuration, with the register qubit(s) in the middle of a 3D cavity, and the agent and environment qubit(s) at opposite sides of it (either longitudinally or transversally to the cavity axis), the number of CNOT gates will not be increased in the actual implementation, i.e., there will be no need for SWAP gates, in the case of two-qubit gates mediated via capacitive coupling. The entangling gates may be implemented in our proposal either via capacitive coupling or via one of the modes of the 3D cavity~\cite{Gate3DCavity}. Currently, experiments with more than 50 entangling gates have been carried out with superconducting circuits~\cite{Barends16}, such that these proof-of-principle quantum reinforcement learning models appear as feasible in this respect. 

Considering a two-qubit gate error of about 0.01, we can estimate the accumulated single-qubit and two-qubit gate error after the protocol to less than 0.05 in all the proposed examples.

\subsubsection*{High-fidelity projective measurement}

The development of Josephson parametric amplifiers\cite{LehnertJPA} allows for projective readout with high fidelity ($\sim99\%$), and similar fidelities have been achieved for two-qubit computational-basis projective-state measurements~\cite{DiCarlo15}, which are the most general ones needed in the proposed protocols. Moreover, repeated quantum nondemolition measurements with this technique provide also values for probabilities that pre- and post-measurement results coincide larger than 0.98\%. In our examples, either one or two subsequent projective measurements are necessary per learning cycle, such that in the first case a measurement error of 0.01 can be assumed, while in the second case a corresponding one of 0.02 is reasonable.

\subsubsection*{Fast closed-loop feedback control }

Finally, the closed-loop feedback control should be much shorter than coherence times in order for the final fidelity to be large. One of the major sources of delay in this process is produced by the controller, as well as the generation and triggering of the microwave pulses to drive the qubits conditionally on the measurement outcomes. Nevertheless, current closed-feedback loop technology with superconducting circuits allows for response times of around 0.11 $\mu$s, which is about two orders of magnitude shorter than the coherence times~\cite{DiCarlo15}. This can be combined with the fast processing provided by a field-programmable-gate-array (FPGA), which also enables more complex signal processing and increases the on-board memory. In our proposed protocols, either one or two projective measurements per learning cycle are considered, and assigning to each of them a feedback loop response time of about 0.11 $\mu$s, there is room for several tens of learning cycles before decoherence starts playing a role.

Summarizing, in the most complex example proposed, the one given in Fig.~\ref{fig:Fig2}, a reasonable estimate of the final fidelity per learning cycle is of 93\%, and the total time of the protocol, assuming a worst-case scenario that the 4 CNOT gates are done sequentially instead of simultaneously, and each takes about 50 ns, would be less than 0.5 $\mu$s. Therefore, there should be room for 4 complete learning cycles of the protocol, which would have a final fidelity about 75\%, assuming that the errors are uncorrelated. 

\section*{Discussion}
We have proposed an implementation of basic protocols in quantum reinforcement learning with superconducting circuits mediated via closed-loop feedback control. Our protocols allow for an agent to acquire information from an environment and modify itself in situ in order to approach the environment state, at the same time as, consequently, modifying it. A possible motivation may be to partially clone quantum information from the environment onto the agent. Current technology has recently enabled all the basic ingredients for carrying out proof-of-principle experiments of quantum reinforcement learning with superconducting circuit platforms, due to improvements in coherence times, gate fidelities, measurements, and feedback loop control. Therefore, it is timely to realize basic quantum machine learning protocols with this quantum implementation, paving the way for future advances in quantum artificial intelligence and its applications.

\section*{Acknowledgements}
The author wishes to acknowledge discussions with U. Alvarez-Rodriguez, M. Sanz, and E. Solano, and support from Ram\'on y Cajal Grant RYC-2012-11391, Spanish MINECO/FEDER FIS2015-69983-P,  Basque Government IT986-16, and UPV/EHU UFI 11/55.

\section*{Author contributions}
L. L. envisioned the project, performed all calculations, analyzed the results, and wrote the manuscript.

\section*{Additional information}

\textbf{Competing financial interests:} The author declares no competing financial interests.

\end{document}